\documentclass[10pt,a4paper]{article}
\pdfoutput=1 %for arxiv
\usepackage[sc]{mathpazo}
\usepackage[compact]{titlesec}

\usepackage{amssymb,amsthm,amsmath,mathtools}
\usepackage{thmtools,thm-restate}
\usepackage[square,numbers]{natbib}%[numbers]
\usepackage{pifont}

\usepackage{tfrupee}
\usepackage{bm}
\usepackage{paralist}
\usepackage[linesnumbered,lined,boxed,ruled,vlined]{algorithm2e}
\DontPrintSemicolon
\SetKwInOut{Parameters}{Parameters}
\SetKwComment{Comment}{$\triangleright$\ }{}
\SetAlFnt{\small}
\SetAlCapFnt{\small}
\SetAlCapNameFnt{\small}
\SetAlCapHSkip{0pt}
\IncMargin{-\parindent}

\usepackage[usenames,svgnames,xcdraw,table]{xcolor}
\definecolor{DarkGreen}{rgb}{0.1,0.5,0.1}
\usepackage[backref=page]{hyperref}
\hypersetup{
	colorlinks=true,
	linkcolor=red,%color of internal links
	urlcolor=DarkGreen,%color of external links
	citecolor=blue%color of links to bibliography
}
\renewcommand*{\backref}[1]{}
\renewcommand*{\backrefalt}[4]{%
    \ifcase #1 (Not cited.)%
    \or        (Cited on page~#2)%
    \else      (Cited on pages~#2)%
    \fi}
\usepackage[capitalise,noabbrev]{cleveref}
\Crefname{property}{Property}{Properties}
\Crefname{theorem}{Theorem}{Theorems}
\Crefname{example}{Example}{Examples}
\Crefname{table}{Table}{Tables}
\Crefname{algorithm}{Algorithm}{Algorithms}

\usepackage{cancel}
\usepackage{tabularx}
\usepackage{nicefrac}
\usepackage{enumerate}
\usepackage{etoolbox}%for \ifstrempty{}
\usepackage[margin=1in]{geometry}
\usepackage[T1]{fontenc}
\usepackage{url}
\usepackage[utf8]{inputenc}
\usepackage{graphicx}
\usepackage{booktabs}
\usepackage{esvect}

\usepackage{mathrsfs,amsfonts,dsfont}
\usepackage{array,multirow,graphicx,bigdelim}

\usepackage{pgf,tikz}
\usepackage{tikz-dimline}
\usepackage{tikz-cd}
\usetikzlibrary{fit,calc,math,shapes,decorations.text,arrows,decorations.markings,decorations.pathmorphing,shapes.geometric,positioning,decorations.pathreplacing}
\tikzset{snake it/.style={decorate, decoration=snake}}

\colorlet{mygray}{gray!40}

\usepackage{relsize}
\newcommand\Tstrut{\rule{0pt}{2.1ex}}       % "top" strut
\newcommand\Bstrut{\rule[-0.9ex]{0pt}{0pt}} % "bottom" strut
\newcommand{\TBstrut}{\Tstrut\Bstrut}

\makeatletter
\let\oldnl\nl% Store \nl in \oldnl
\newcommand{\nonl}{\renewcommand{\nl}{\let\nl\oldnl}}% Remove line number for one line
\makeatother

  {\list{}{\leftmargin=#1\rightmargin=#1}\item[]}%
  {\endlist}

\usepackage[labelfont={normalfont,bf},textfont=it]{caption}
\usepackage{subcaption}

\theoremstyle{definition}

\theoremstyle{remark}

\newcommand{\weat}{\text{\bf \textup{\fontfamily{cmss}\selectfont WEAT}}}

\newcommand{\ik}{\text{\bf {\fontfamily{cmss}\selectfont IndK}}}
\newcommand{\mean}{{\textup{mean}}}
\newcommand{\stddev}{{\textup{stddev}}}
\newcommand{\maleper}{{w_{\textup{MP}}}}
\newcommand{\femaleper}{{w_{\textup{FP}}}}
\newcommand{\malevic}{{w_{\textup{MV}}}}
\newcommand{\femalevic}{{w_{\textup{FV}}}}
\usepackage{xspace}
\renewcommand{\AA}{\ensuremath{\mathcal A}\xspace}
\newcommand{\BB}{\ensuremath{\mathcal B}\xspace}

\newcommand{\DD}{\ensuremath{\mathcal D}\xspace}

\newcommand{\XX}{\ensuremath{\mathcal X}\xspace}
\newcommand{\YY}{\ensuremath{\mathcal Y}\xspace}

\usepackage{flushend}
\usepackage[utf8]{inputenc}
\usepackage[inline]{enumitem}
\Crefname{claim}{Claim}{Claims}

\allowdisplaybreaks

\providecommand{\keywords}[1]{\textbf{\textit{Keywords:}} #1}

\title{Disentangling Societal Inequality from Model Biases: Gender Inequality in Divorce Court Proceedings}
\author{Sujan Dutta\\
  {Rochester Institute of Technology}\\
  \texttt{sd2516@rit.edu}
  \and
  Parth Srivastava\\
  {Indian Institute of Technology, Kanpur}\\
  \texttt{parthsri@iitk.ac.in} \\
\and
Vaishnavi Solunke \\
  {Rochester Institute of Technology}\\
  \texttt{vs5709@rit.edu} \\
 \and
Swaprava Nath\\
  {Indian Institute of Technology, Bombay}\\
  \texttt{swaprava@cse.iitb.ac.in}\\ 
  \and
  Ashiqur R. KhudaBukhsh~\thanks{This work is accepted at IJCAI 2023 (AI for good track). Ashiqur R. KhudaBukhsh is the corresponding author.}\\
  {Rochester Institute of Technology}\\
  \texttt{axkvse@rit.edu}}
\date{}

\begin{document}

\maketitle

\begin{abstract}
 Divorce is the legal dissolution of a marriage by a court. Since this is usually an unpleasant outcome of a marital union, each party may have reasons to call the decision to quit which is generally documented in detail in the court proceedings. Via a substantial corpus of 17,306 court proceedings, this paper investigates gender inequality through the lens of divorce court proceedings. While emerging data sources (e.g., public court records) on sensitive societal issues hold promise in aiding social science research, biases present in cutting-edge natural language processing (NLP) methods may interfere with or affect such studies. We thus require a thorough analysis of potential gaps and limitations present in extant NLP resources. In this paper, on the methodological side,  we demonstrate that existing NLP resources required several non-trivial modifications to quantify societal inequalities. On the substantive side, we find that while a large number of court cases perhaps suggest changing norms in India where women are increasingly challenging patriarchy, AI-powered analyses of these court proceedings indicate striking gender inequality with women often subjected to domestic violence.
\end{abstract}


\section{Introduction}
\label{sec:intro}

The 2011 decennial census in India gave its citizens the following choices to select their marital status -- never married, separated, divorced, widowed, married. Based on the census data, a study reported some startling facts~\cite{jacob2016marriage}: 1.36 million of the Indian population is divorced which accounts for 0.24\% of the married population, and 0.11\% of the total population. More women were separated or divorced than men, and the number of separation was almost three times as high as the number of divorce. 

Divorce, a historically taboo topic in India for ages~\cite{dommaraju2016divorce}, seldom features in mainstream Indian discourse~\cite{goode1962marital}. Recent indications of changing social acceptance of divorcees notwithstanding~\cite{mani2017study}, divorce in India still carries a considerable social stigma~\cite{belliappa2013gender}.  

\emph{How do we quantify gender inequality in Indian divorce?} Surveys about divorce often have limited participation and a small sample size~\cite{vasudevan2015causes}, perhaps due to the social stigma attached. A vulnerable community -- Indian women under conjugal distress -- had limited visibility to social scientists. Via a substantial corpus of 17,306 divorce court proceedings, this paper conducts the first-ever computational analysis of gender inequality in Indian divorce based on public court records.

\begin{figure}[h!]
\centering
\fbox{\includegraphics[trim={0 0 0 0},clip, height=1.0in]{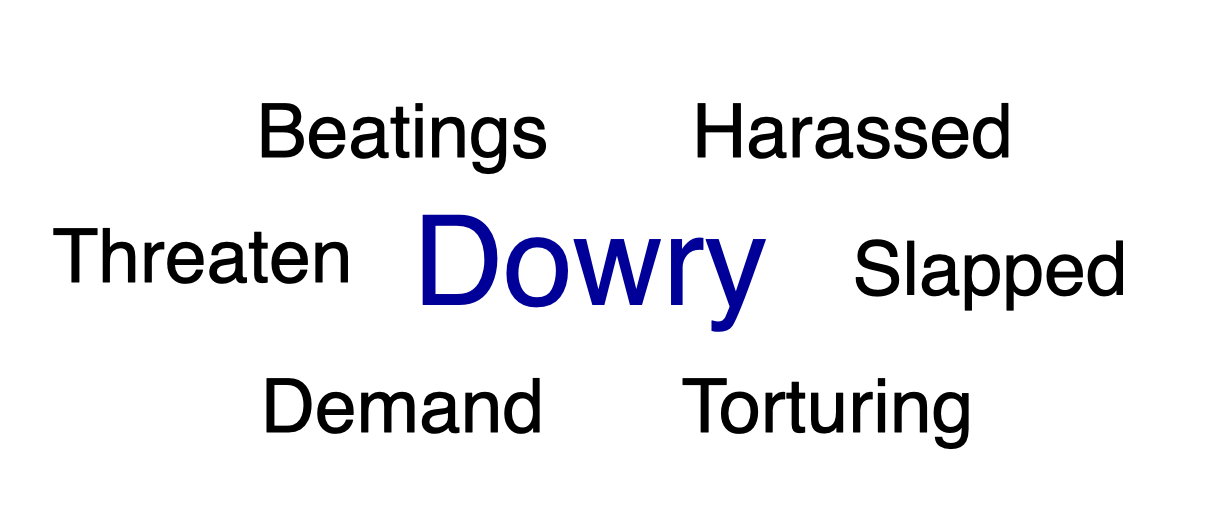}}
\caption{Nearest neighbors of the word \texttt{dowry} in a word embedding trained on divorce court proceedings from Rajasthan, a state from India. Despite legal prohibition since 1961~\cite{rao1973dowry}, this retrogade social practice has continued in India with several studies linking it to other social crises such as female feticide, domestic abuse and violence, and dowry deaths~\cite{ahmad2008dowry}. } 

\label{fig:dowry-cloud}
\end{figure}

Even though written in English, legal texts are often domain-specific~\cite {bhattacharya2019comparative}. The considerable variation of legal jargon across countries and courts makes domain-specific analysis important. In that vein, Indian legal NLP is an emerging field~\cite{bhattacharya2019comparative,kalia2022classifying}. Most NLP research on legal texts thus far has focused on building robust tools to analyze legal text. Recent research, however, on in-group bias~\cite{ash2021group} and sexual harassment~\cite{kumar2020sexual}, and \Cref{fig:dowry-cloud} and \Cref{tab:example} suggest that automated methods to glean social insights from large-scale, legal texts merit investigation. Barring few recent lines of work~\cite{madaan2018analyze,DBLP:conf/acl-trac/BhattacharyaSKB20,khadilkar2021gender}, there is surprisingly little literature on large-scale linguistic analysis of gender bias in India, let alone on legal text zeroing in on divorce.  

\renewcommand{\tabcolsep}{2mm}
\begin{table*}[t]
{
\small
\begin{center}
     \begin{tabular}{|p{0.45\textwidth}|p{0.45\textwidth}|}
  \hline
  \textit{husband} $\rightarrow$ \textit{wife} &  \textit{wife} $\rightarrow$ \textit{husband} \\
    \hline

\cellcolor{blue!15}In the petition, it is alleged that the respondent always mentally and physically harassed the petitioner and threatened the petitioner that she will commit suicide and sometimes, she even physically tortured the petitioner and she used to beat and slap the petitioner when she becomes angry. 
& 
\cellcolor{red!15}
The respondent is a drunkard and he used to consume alcohol everyday and he ill treated and tortured the petitioner, demanding more dowry. \\ \hline
\cellcolor{blue!15}
It was held that a husband cannot be expected to continue to live with the wife in the face of her sustained attitude of causing humiliation and calculated torture.
&
\cellcolor{red!15}
The interaction of the petitioner with her friends and relatives were viewed with suspicion. The petitioner's telephone facility was disconnected. The Email account of the petitioner was checked by the respondent and often her friends and colleagues were threatened and abused.
\\ \hline
\cellcolor{blue!15}
According to him, he was working abroad till 2019 and only when he returned to India did he realise that he was cheated.
&
\cellcolor{red!15}
After considering the evidence, it was found that she was not willing to go with her husband further as he had cheated her once.
\\ \hline

    \end{tabular}
    
\end{center}
\caption{{Snippets from court proceedings where unpleasant verbs (e.g., cheat, torture, slap, abuse, beat, threaten etc.) are used. The left column ($\textit{husband} \rightarrow \textit{wife}$) accuses the wife of the wrongdoing. The right column ($\textit{wife} \rightarrow \textit{husband}$) accuses the husband of the wrongdoing.}}
\label{tab:example}}

\end{table*}

While emerging data sources (e.g., public court records available on the web) offer opportunities for social scientists to study important and sensitive social issues that previously had limited survey data, applying cutting-edge NLP methods to newer domains requires careful evaluation of the critical question: \textit{How much of the (perceived) gender inequality as quantified by the methods truly reflects the corpus and how much of it is due to the inherent biases of the employed NLP methods?} 
In this paper, we show that the subtleties present in legal text present unique challenges. Unless we consider them and make non-trivial changes to existing methods, we may end up drawing inaccurate social conclusions. 
We further show that sophisticated NLP methods built on top of large language models (LLMs) need scrutiny when applied to social inference tasks involving genders. We, in fact, conduct a much broader \textit{bias audit} of these systems. Our audit reveals well-known LLMs often exhibit gender bias even on simple subject-verb-object sentence completion tasks. Through a corpus-specific text entailment analysis, we demonstrate that downstream applications such as natural language inference (NLI) systems also exhibit sensitivity to gender. We finally, present a novel inconsistency sampling method to mitigate this bias and present our social findings.     

\smallskip \noindent
To summarize, our contributions are the following:
\paragraph{Social:} We create a substantial corpus of 17,306 divorce court proceedings and conduct the first-ever analysis of gender inequality through the lens of divorce proceedings. While a large number of court cases perhaps suggest changing norms in India where women are increasingly challenging patriarchy~\cite{sonawat2001understanding}, our analyses reveal widespread domestic violence, dowry demands, and torture of the bride.

\paragraph{Methodological:} We address extant gaps and limitations in multiple NLP frameworks. We propose non-trivial modifications to the \weat{} framework~\cite{caliskan2017semantics} to make it suitable for  legal text.
We demonstrate a novel application of text entailment~\cite{maccartney2008modeling} in quantifying gender inequality. We investigate several potential sources for model bias in NLP resources that can interfere with quantifying gender inequality. We present a novel inconsistency sampling method exploiting counterfactuals to mitigate this bias.

\section{Dataset}

\subsection{Collection}
We scrape all the publicly available court proceedings with the word \texttt{divorce} between January 1, 2012 to December 31, 2021 from \url{https://indiankanoon.org/} (hereafter \ik), an Indian law search engine launched in 2008 and the largest free online repository of the court proceedings of different courts and tribunals of India. Prior computational law research ~\cite{mandal2021unsupervised} and gender focused social science studies~\cite{kumar2020sexual} have used \ik\ as source of data. 

We download 86,911 case proceedings containing the word \texttt{divorce} from \ik\ using its advanced search feature. Filtering based on the keyword \texttt{divorce} is a high-recall approach to obtain relevant cases with precedence in computational social science research~\cite{HaltermanKSO21Policing,DuttaPolice}. However, the presence of the keyword \texttt{divorce} may not always indicate a divorce court proceeding; for instance, the keyword can be used to describe the marital status of any of the litigants. It can also be used in an altogether different context (e.g., \emph{divorced from reality}). We use the following heuristic to further refine our dataset. We also look for other words (e.g., \texttt{husband}, \texttt{wife}, \texttt{marriage}) and phrases (e.g., \texttt{decree of divorce}), and check if such occurrences repeat for a minimum threshold (set to 5). On a random sample of 100 cases after we apply this cleaning method, a manual inspection reveals that 94 are divorce cases. Hence, our keyword-based filtering is reasonably precise. This pruning step retains 25,635 cases.  

\subsection{Data Pre-processing}

To quantify gender inequality in court proceedings, we must disambiguate the legal parties -- the plaintiff and the defendant -- and accurately identify of the husband and the wife, who plays which role. Indian legal documents use a wide range of legal terms to denote the plaintiff (e.g., appellant, applicant, complainant, petitioner) and the defendant (e.g., respondent, nonapplicant, opponent). We observe different courts have different formats (sometimes, multiple formats) to summarize the proceedings.  The documents also specify which party in marriage represents which role in several different ways (e.g., respondent/wife, respondent-wife, respondent aggrieved wife). We write a regular-expression-based pipeline and consolidate such information to identify the gender of the plaintiff and the defendant across all the states.

The names and salutations (e.g., \texttt{Mr.}, \texttt{Mrs.}, \texttt{Smt.}, \texttt{Shri}) of the plaintiff and defendant also provide gender information. Subcultural naming conventions played a key role in assigning gender to the litigants in some of the cases. For instance, \texttt{Kaur}, meaning princess, is a Punjabi last name only for females~\cite{kaur2019gap}. Or \texttt{ben}, meaning sister, is solely used in many female names in Gujarat~\cite{mistry1982personal}. Dependence information of the litigants also provides gender information (e.g., \texttt{son of}, \texttt{daughter of}, \texttt{wife of}).\footnote{We did not find a single mention of \texttt{husband of} in our dataset.}

Of the 25,635 cases, we could unambiguously assign gender to 17,306 cases. For each case, we replace each mention of the litigants as \texttt{wife} or \texttt{husband} accordingly. For example, a proceeding snippet ``\textit{The \colorbox{blue!25}{plaintiff/wife} has filed for a divorce. The \colorbox{blue!25}{plaintiff} was married to the \colorbox{red!25}{defendant} for three years.''}, will be modified to ``\textit{The \colorbox{blue!25}{wife} has filed for a divorce. The \colorbox{blue!25}{wife} was married to the \colorbox{red!25}{husband} for three years.}'' This data set, $\DD_\textit{divorce}$, consists of 30,615,754 (30 million) tokens.

\section{Brief overview of Indian legal system}

Indian Judicial System is largely based on the English Common Law system (where, the law is developed by judges through their decisions, orders, and judgments). The nation has 28 states and 8 union territories (UT), and a total of 25 high courts (some high courts have jurisdiction of more than a state or UT). The federal structure has a supreme court coupled with the high courts that roughly handle the cases in a state or UT. The legal cases of divorce are usually handled by the family or district courts. However, some unresolved cases or sometimes fresh cases are also heard by the high courts. Since the court proceedings are public records and are digitally made available freely by \ik, we found this dataset to be quite appropriate for a large-scale study on gender equality in court proceedings. 

\begin{figure}[t!]
     \centering
     \begin{subfigure}[b]{0.45\textwidth}
         \centering
            \includegraphics[width = \textwidth]{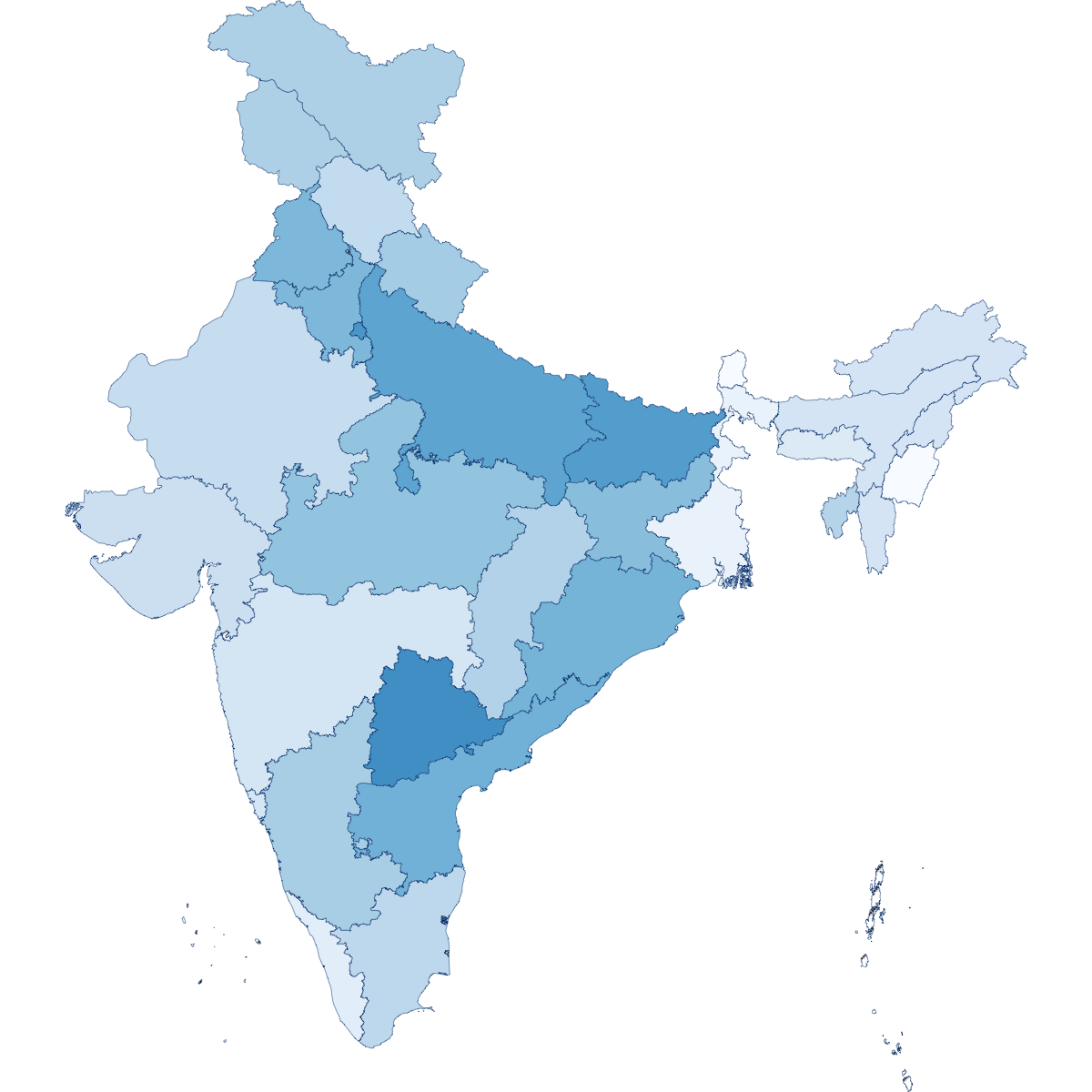}
         \caption{Dowry}
         \label{fig:dowry}
     \end{subfigure}
     \hfill
     \begin{subfigure}[b]{0.45\textwidth}
         \centering
            \includegraphics[width = \textwidth]{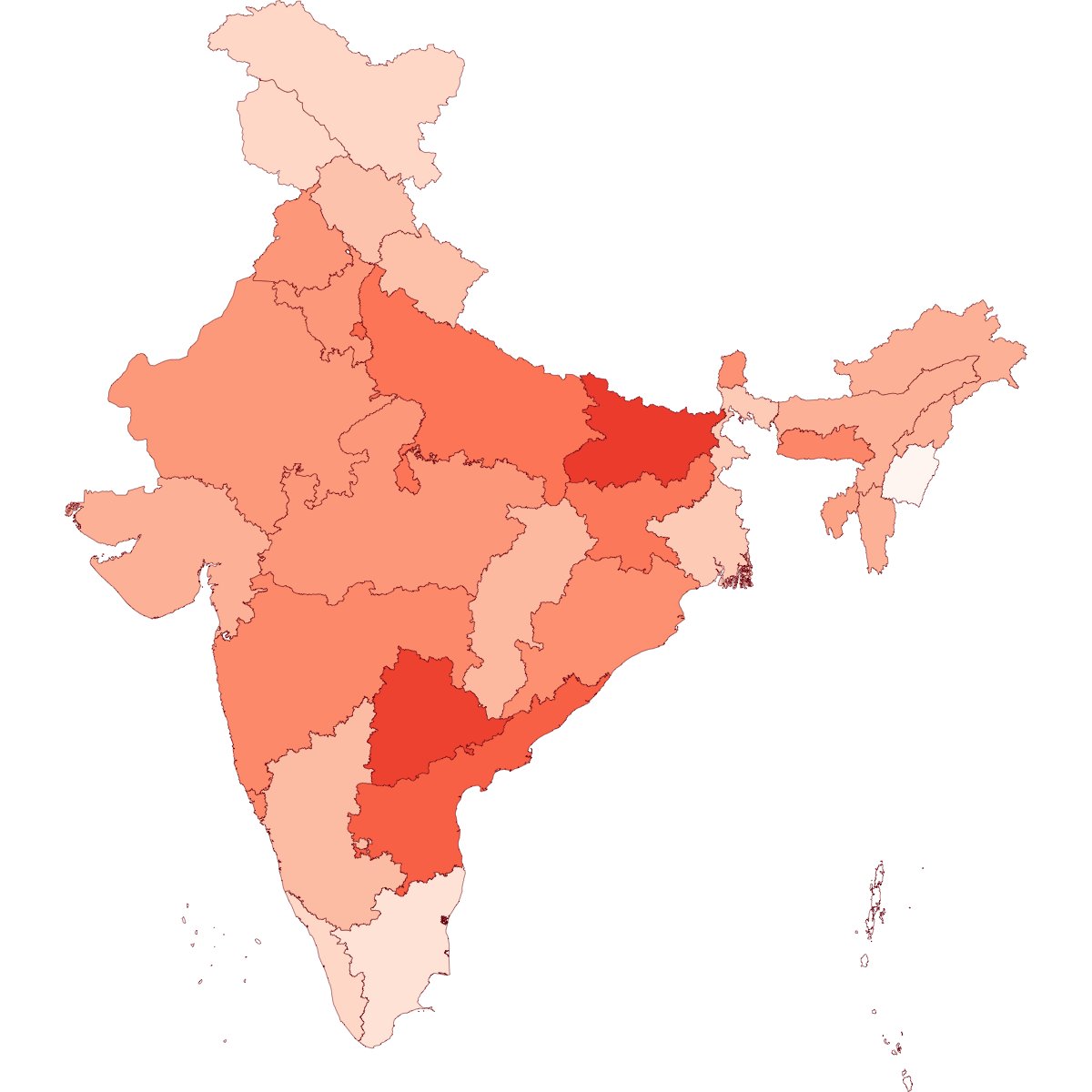}
         \caption{IPC 498-A}
         \label{fig:498}
     \end{subfigure}
        \caption{Choropleths of divorce cases mentioning \texttt{dowry} and \texttt{IPC 498-A}. For each state, we compute the total number of divorce cases that mention \texttt{dowry} (for Figure~\ref{fig:dowry}) or \texttt{498-A} (for Figure~\ref{fig:498}) at least once and divide by the total number of divorce cases in that state. Each number is in the range [0, 1]. A larger number indicates greater mention of \texttt{dowry} or \texttt{498-A}.  Higher intensity colors indicate larger values. Section 498-A is a section in the Indian Penal Code (IPC) introduced in 1983 to protect women from \textit{marital cruelty}. The base maps used for this plot are sourced from the Government of India. The authors are aware that these maps include disputed territories. These maps do not constitute judgments on existing disputes.}
        \label{fig:DowryIPC}
\end{figure}

\section{Dowry in Divorce Proceedings}

The dowry system involves a transaction of financial assets between the bride's family and the bridegroom's family with the latter being the recipient of the financial assets. While legally prohibited in India since 1961~\cite{rao1973dowry}, this practice has continued well after its legal prohibition and has a strong link to social crises such as female feticide~\cite{ghansham2002female}, domestic abuse and violence~\cite{banerjee2014dowry,rastogi2006dowry}, and dowry deaths~\cite{ahmad2008dowry}. In order to protect the bride from marital cruelty  and domestic violence, Indian Penal Code introduced Section 498 in 1983~\cite{carpenter2016protecting}. 

Figure~\ref{fig:DowryIPC} reflects relative proportions of divorce cases containing the text tokens \texttt{dowry} and \texttt{498-A}. For each state, we report the fraction of divorce cases that contain at least one mention of these two tokens. A higher intensity color indicates a larger proportion of such cases. We observe that overall, 24.38\% of all cases and 21.86\% of all cases mention \texttt{dowry} and \texttt{498-A}, respectively.
Jacob and Chattopadhyay, \cite{jacob2016marriage} reported that divorce in India does not follow any one-size-fits-all pattern across different states; there exists sufficient interstate variation even for the rate of divorce. We notice a considerable variation in mentions of dowry and section 498-A across different states indicating variance in reported cases of dowry or domestic violence. Among the states and the union territories, the top three entries in terms of dowry mentions are Telangana, Delhi, and Bihar while the top three entries in terms of Section 498-A mentions are Bihar, Telangana, and Andhra Pradesh. Bihar and Telangana have social science literature documenting dowry and domestic violence~\cite{babu2011dowry,jakimow2013everyone}.  Apart from the overlap in the top three entries, the statewise dowry and 498-A mentions are moderately correlated (correlation coefficient: 0.67). 

We next conduct a qualitative analysis of (alleged) dowry demands~\footnote{This analysis follows the statements made by the plaintiffs}. On a random sample of 100 court proceedings where the (alleged) dowry demand is explicitly recorded, we observe that the estimated demanded amount is \rupee393,100 $\pm$ 544,876. We observe demanded amounts as low as \rupee5,000 to as high as \rupee3,000,000 which explains the staggeringly high variance in our estimation. This also indicates the broad economic spectrum present in India and how far and wide the system of dowry (allegedly) persists. We further observe that cash is not always the solely demanded financial asset. Gold is the second-most commonly demanded asset. Out of the 100 cases, 34 cases report gold demands (71.2 $\pm$ 84.6 gm). When we adjust the valuation of demanded gold replacing it with the historical average gold price in India across 2012 and 2021~\footnote{Obtained from \url{https://www.bankbazaar.com/gold-rate/gold-rate-trend-in-india.html}}, the estimated (alleged) demanded dowry is \rupee474,798 $\pm$ 567,219.

\section{Methods Overview}

We use two NLP methods to quantify gender inequality: (1) Word Embedding Association Test; and (2) a text entailment framework. A brief description  follows. 

\subsection{Word Embedding Based Methods}

The first metric is called \texttt{\bf W}ord \texttt{\bf E}mbedding \texttt{\bf A}ssociation \texttt{\bf T}est (\weat) introduced by \cite{caliskan2017semantics}. To calculate the metric, the words are embedded and the vectors $\vv{\bm{a}}$ and $\vv{\bm{b}}$ are obtained for the words $a$ and $b$ respectively. The cosine similarity of these words are denoted by $\cos(\vv{\bm{a}},\vv{\bm{b}})$. The metric considers two sets of {\em target words} given by $\XX$ and $\YY$, and two sets of {\em attribute words} $\AA$ and $\BB$. Then, the \weat\ score is defined as $\weat{}(\XX, \YY, \AA, \BB) = (\mean_{x \in \XX} \sigma(x, \AA, \BB) - \mean_{y \in \YY} \sigma(y, \AA, \BB))/\stddev_{w \in \XX \cup \YY} \sigma(w, \AA, \BB),$
% \begin{gather}
% \label{eqn:weat}
%     \weat{}(\XX, \YY, \AA, \BB) = \frac{\mean_{x \in \XX} \sigma(x, \AA, \BB) - \mean_{y \in \YY} \sigma(y, \AA, \BB)}{\stddev_{w \in \XX \cup \YY} \sigma(w, \AA, \BB)},
% \end{gather}
where, $\sigma(w, \AA, \BB) = \mean_{a \in \AA} \cos(\vv{\bm{w}},\vv{\bm{a}}) - \mean_{b \in \BB} \cos(\vv{\bm{w}},\vv{\bm{b}}).$
% \begin{gather*}
%  \sigma(w, \AA, \BB) = \mean_{a \in \AA} \cos(\vv{\bm{w}},\vv{\bm{a}}) - \mean_{b \in \BB} \cos(\vv{\bm{w}},\vv{\bm{b}}).
% \end{gather*}
Intuitively, $\sigma(w, \AA, \BB)$ measures the association of $w$ with the attribute sets, and the \weat{} score measures the differential association of the two sets of target words with the attribute sets. A positive \weat{} score implies that the target words in $\XX$ is more associated with the attribute words in $\AA$ than $\BB$ and the words in $\YY$ is more associated with $\BB$ than $\AA$.

\subsection{Text Entailment Based Methods}

Quantifying gender inequality relying on the distributed representation of words presents a diffused, bird's-eye view of the larger trends. Also, these methods are known to be data-hungry~\cite{DBLP:conf/nips/MikolovSCCD13}. Data availability often becomes a limiting factor to conducting contrastive studies at different spatio-temporal granularity. In what follows, we present a novel application of text entailment, a natural language inference (NLI) task~\cite{dagan2005pascal} that bypasses the data size requirement and equips us with a finer lens through which we can compare and contrast gender inequality with respect to individual verbs.

An NLI system take a premise $\mathcal{P}$ and a hypothesis $\mathcal{H}$ as input and outputs entailment, contradiction, or  semantic irrelevance. For instance, the hypothesis \emph{some men are playing a sport} is entailed by the premise \emph{a soccer game with multiple males playing}~\cite{bowman-etal-2015-large}. As one can see, textual entailment is more relaxed than pure logical entailment and it can be viewed as a human reading $\mathcal{P}$ would infer most likely $\mathcal{H}$ is true. This framework has gained traction in several recent social inference tasks that include estimating media stance on policing~\cite{halterman-etal-2021-corpus,DuttaPolice}, aggregating social media opinion on election fairness~\cite{Capitol2022}, and detecting COVID-19 misinformation~\cite{hossain-etal-2020-covidlies}.
Formally, let \textit{NLI}($\mathcal{P}$,$\mathcal{H}$) takes a premise $\mathcal{P}$ and a hypothesis $\mathcal{H}$ as inputs and outputs $o \in \{\textit{entailment}, \textit{contradiction}, \textit{neutral}\}$. Following~\cite{DuttaPolice}, we define entailment ratio (denoted by \textit{ent}($\mathcal{D}$, $\mathcal{H}$)) for given corpus $\mathcal{D}$ and a hypothesis $\mathcal{H}$, as the fraction of the individual sentences present in $\mathcal{D}$ that entails $\mathcal{H}$:
\textit{ent}($\mathcal{D}, \mathcal{H}$) =  $\frac{\sum_{\mathcal{P} \in \mathcal{D}} \mathds{I}(\emph{NLI}(\mathcal{P}, \mathcal{H}) =  \textit{entailment})}{|\mathcal{D}|}$,
where $\mathds{I}$ is the indicator function. A larger value of \textit{ent}($\mathcal{D}, \mathcal{H}$) indicates greater support for $\mathcal{H}$ in the corpus.

Consider we are interested in learning how often the husband and the wife are accused of torture (physical or emotional) in our corpus. We analyze this research question in the following way. We first construct a sub-corpus $\mathcal{D}_\textit{torture}$ from the divorce court proceedings consisting of sentences that (1) mention \texttt{husband} or \texttt{wife} at least once; and (2) mention \texttt{torture} as a verb at least once. We next construct two hypotheses -- $\mathcal{H}_{\texttt{MV},torture}$ and $\mathcal{H}_{\texttt{FV},torture}$ -- using a \texttt{man} and a \texttt{woman} as victims and perpetrators interchangeably.  $\mathcal{H}_{\texttt{MV},\textit{torture}}$ is \emph{A \colorbox{blue!25}{woman} tortures a \colorbox{red!25}{man}} and $\mathcal{H}_{\texttt{FV},\textit{torture}}$ is \emph{A \colorbox{red!25}{man} tortures a \colorbox{blue!25}{woman}}. We next compute the entailment gap defined as\\$
\textit{gap}(\mathcal{D}_\textit{torture},\textit{torture}) = \\\textit{ent}(\mathcal{D}_\textit{torture},\mathcal{H}_{\texttt{FV},\textit{torture}}) - \textit{ent}(\mathcal{D}_\textit{torture},\mathcal{H}_{\texttt{MV},\textit{torture}})
$\\
Effectively, this means we compute the fraction of sentences that entail \emph{A \colorbox{blue!25}{woman} tortures a \colorbox{red!25}{man}} in $\mathcal{D}_\textit{torture}$  and subtract it from the fraction of sentences that entail \emph{A \colorbox{red!25}{man} tortures a \colorbox{blue!25}{woman}} in $\mathcal{D}_\textit{torture}$. An overall positive number indicates that the male has been described as the torturer more often than the female in court proceedings. A negative value would indicate the opposite way. Similar analysis can be extended to other verbs such as \texttt{assault}, \texttt{beat}, or \texttt{abuse}.

\section{Design Considerations}

Adapting the \texttt{WEAT} and entailment frameworks to quantify gender inequality in our domain requires careful consideration of several aspects described in what follows. 

\subsection{Verbs for Target Sets}

Traditionally, \weat{} score is used to quantify gender or racial stereotypes. Majority of the elements present in those attribute sets would be nouns and adjectives (e.g., criminals, terrorists, doctors, police)~\cite{DBLP:conf/naacl/ManziniLBT19,greenwald2014malice} and seldom verbs~\cite{DBLP:conf/acl/HoyleWWAC19}. 
We are interested in understanding the action space of the two parties fighting a divorce case; we want to know if the court described that one party tortured or abused the other. Hence, verbs are a natural choice for our target set.  

We inspect the list of high-frequency verbs in the corpus and narrow down to the following ten verbs: $\XX_\textit{unpleasant}$ = 
\{\texttt{abuse},
\texttt{assault},
\texttt{beat}, 
\texttt{burn}, 
\texttt{cheat},
\texttt{misbehave},
\texttt{rape}, 
\texttt{slap}, \texttt{threaten}, \texttt{torture}\}. A small subset of these words are already present in the list of unpleasant stimuli presented in~\cite{greenwald2014malice}. We further compute the average valence score of these words as per the lexicon presented in~\cite{warriner2013norms}. We find the average valence score of $\XX_\textit{unpleasant}$ is 2.7, comparable to the average valence score (2.16) of unpleasant stimuli presented in~\cite{greenwald2014malice}.  

Divorce being a bitterly fought family situation, we observe a sparse presence of pleasant verbs such as \texttt{love}, \texttt{care}, or \texttt{empathy} in our corpus. Since infrequent words in the corpus do not have reliable embeddings~\cite{DBLP:conf/iclr/LampleCRDJ18}, in contrast with traditional applications of \weat{} score, we choose the target set $\YY$ to be an empty set. 

\subsection{The Torturer and the Tortured}

The attribute sets $\AA$ and $\BB$ as defined in the \weat\ score represents the identifiers used for the plaintiff and defendant in our data (e.g., $\AA$ consisting of \texttt{he}, \texttt{him}, \texttt{husband}, and $\BB$ consisting of \texttt{she}, \texttt{her}, \texttt{wife} etc.). However, notice that \weat\ score is agnostic about whether the identifier is the {\em contributor} or the {\em receptor} of target words. For example, torture does not happen in isolation; it requires a torturer and one who is tortured. Unlike nouns, verbs are typically associated with two entities -- the subject and the object. To disambiguate between ``\textit{the \colorbox{red!25}{husband} tortured the \colorbox{blue!25}{wife}}'' and ``\textit{the \colorbox{blue!25}{wife} tortured the \colorbox{red!25}{husband}}'', a word embedding needs to understand this nuance. Otherwise, the embedding is likely to place both the plaintiff and defendant identifiers equidistant to the verb. 

%To verify this, we constructed two extreme datasets where the first one has one million repetitions of the sentence ``\texttt{He regularly tortured her}'' and ten thousand repetitions of ``\texttt{She regularly tortured him}'', and the second one has the same two sentences with the number of occurrences interchanged. Unsurprisingly, we find that the \weat{} score of these two corpora are similar:  0.003\sn{TODO} for the first dataset and 0.02\sn{TODO} for the second.

To disambiguate these two situations, we run the corpus through the \texttt{stanza} POS tagger \cite{qi2020stanza} to find out the subject and object of the sentences and whether the statements are in active or passive voice. Based on this, we classify the subjects and objects as `male perpetrator', `female perpetrator', `male victim', or `female victim', in the sentences that has the target verbs. We replace these four cases with four unique words (denoted by $\maleper, \femaleper, \malevic$, and $\femalevic$, respectively) so that those words do not occur anywhere else in any of the documents. We call this new dataset $\DD_\textit{replaced}$.

\section{Word Embedding Based Analysis}

We are interested in two research questions: 

\noindent\textit{\textbf{RQ 1:} How does gender inequality manifest in divorce court proceedings with respect to unpleasant verbs in $\mathcal{X}$?}
\\
\textit{\textbf{RQ 2:} Is our careful disambiguation of the torturer and the tortured necessary at all?}

\begin{figure}[h!]
\centering
    \includegraphics[width=0.5\linewidth]{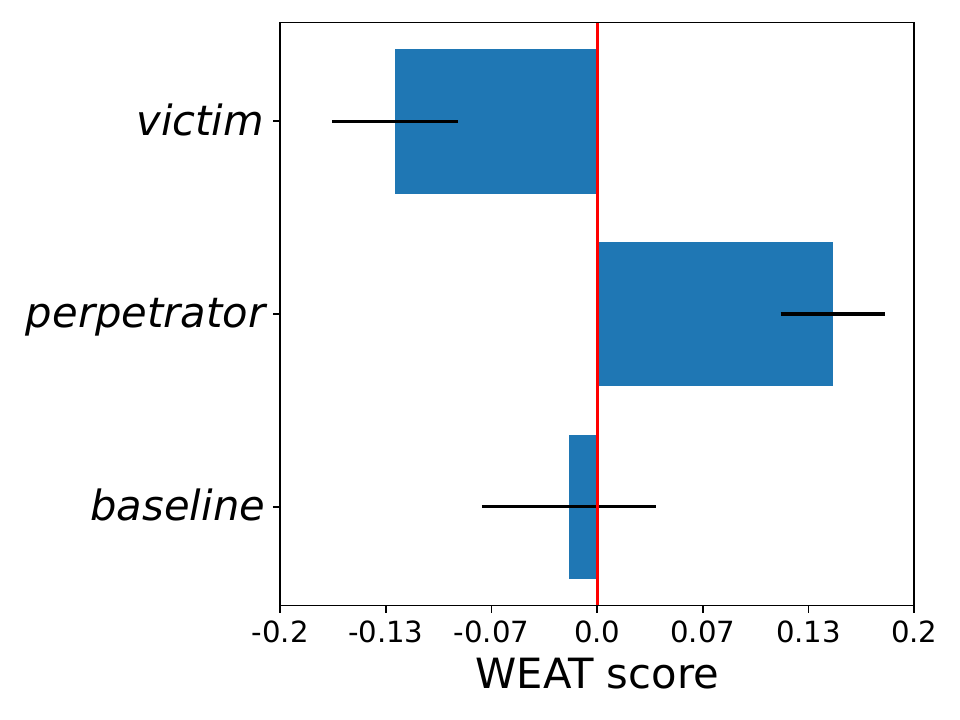}
    \caption{\weat{} scores.  The \weat{} score is averaged over ten runs. A larger positive value indicates a greater bias toward men. The top two values (victim and perpetrator) are computed on $\mathcal{D}_\textit{replaced}$. The bottom (baseline) value is computed on $\mathcal{D}_\textit{divorce}$. The top value (victim) is from the perspective of the victim where the attribute sets \AA and \BB are set to words denoting male victims and female victims, respectively. A negative value implies that women are more associated with the unpleasant verbs in $\mathcal{X}_\textit{unpleasant}$. The middle value (perpetrator) is from the perspective of the perpetrator. A positive value implies that men are more associated with the unpleasant verbs in $\mathcal{X}_\textit{unpleasant}$. The baseline indicates that without incorporating this nuance, the \weat{} framework will present an inaccurate evaluation of the social bias present in the corpus.}
    \label{fig:weat}
\end{figure}

In order to answer these two questions, we run two sets of experiments with identical training configurations. First, we run experiments on $\DD_\textit{replaced}$ using the target and attribute sets as defined in the previous section. We train the word embedding model 10 times and calculate the \weat{} scores for each of the following two cases: when both genders are (a)~perpetrators, i.e., when $\AA=\{ \maleper \}, \BB=\{ \femaleper \}$, and (b)~victims, i.e., when $\AA=\{ \malevic \}, \BB=\{ \femalevic \}$. We use the default parameters for training our {\tt FastText}~\cite{bojanowski2017enriching} Skip-gram embedding with the dimension set to 100 for all word-embeddings in this paper. Second, we run a {\em baseline} experiment with the original text data without replacing them with the four unique words ($\DD_\textit{divorce}$) and use the attribute sets as $\AA=\{ \texttt{husband} \}$ and $\BB=\{ \texttt{wife} \}$. The number of runs and the embedding method are the same in both experiments. The results are shown in \Cref{fig:weat}.

%Hyperparameters are listed in the SI. An additional experiment with a different choices of $\mathcal{A}$ and $\mathcal{B}$ is summarized in the SI.

As already described, a negative \weat{} score indicates $\BB$ is more associated with the target set as compared to $\AA$. Hence, if we look from the perspective of the victim, we find that women are more associated with the unpleasant verbs than men. In contrast, when viewed from the perpetrator's perspective, a positive \weat{} score implies that men are more associated with the unpleasant verbs. Hence, our results indicate that in our corpus, women are more often the victims while men are more often the perpetrators. 

Our baseline experiments that do not make any distinction between the perpetrator and the victim give a \weat{} score close to zero indicating near-perfect gender equality. This inaccurate result, while highly surprising from a social science perspective, is not unexpected given how the original framework functions. The two entities (husband and wife) are present around the unpleasant verbs with nearly equal frequency. If the method does not make any distinction between the roles of victim and perpetrator, \weat{} will give inaccurate results. We thus carefully use the \weat{} score to elicit the correct
gender bias when applied to legal texts for our social science research question. 

%In the interest of space, we present  granular analyses with \ripa{} scores in the SI. 

\section{Societal Inequality and Model Bias}

Our word embeddings are computed from scratch while our next set of experiments relies on downstream applications built on top of large language models. Large language models (LLMs) are known to have a wide range of biases due to the train data~\cite{bender2021dangers} and extant literature has examined gender bias in the form of occupational stereotypes present in NLI systems~\cite{rudinger2017social}. We thus need to disentangle societal inequalities that are potentially reflected in our corpus and model biases that are potentially present in the NLP applications. 

Essentially, for a premise/hypothesis pair $\langle\mathcal{P},\mathcal{H}\rangle$, the NLI system estimates the probability P($\mathcal{H}$ $|\mathcal{P}$). However, how LLMs encode the probability P($\mathcal{H}$) when the hypotheses primarily consists of the two genders (male and female) and a set of verbs is  understudied. A thorough investigation first reveals that the masked word prediction probability of several well-known LLMs is sensitive to gender. We next present a measure to quantify gender bias sensitivity of NLI frameworks and present mitigating strategies. Finally, we use a bias-mitigated NLI system on our corpus and report findings. 

\subsection{Implicit Bias in Agent and Theme in LLMs}

Unlike existing literature that primarily target occupational stereotypes to quantify and analyze gender bias~\cite{rudinger2017social,DBLP:journals/corr/abs-2105-05541,khadilkar2021gender,caliskan2017semantics,kumar2020nurse}, we focus on a very basic unit in a sentence -- the verbs. Following~\cite{SAPPowerAgency}, let in a sentence \textit{X verbs Y}, \textit{X} represent the agent and \textit{Y} represent the theme. 
Many verbs imply the relative authority levels between the agent and the theme. For example, in the sentence \textit{The football coach instructed the players to play a conservative game}, the agent (the football coach) has more authority than the theme (the players). In contrast, the agent has less authority than the theme in the sentence \textit{The football coach honored the players' suggestion to play a conservative game}. First proposed in~\cite{SAPPowerAgency}, the connotation relation of power captures this notion of power differential between an agent and a theme with respect to a given verb.  

While the connotation relation of power has been analyzed in the context of gender inequality in movie scripts~\cite{SAPPowerAgency} and follow-on research focused on editorial fixes to remove bias~\cite{PowerTransformer}, little or no literature exists that documents the implicit gender bias present towards the agent and the theme when specific verbs are considered. This research is important and has a broader impact beyond our current social inference task. For instance, if an LLM encodes that it is less likely for a woman to inspire or guide someone than a man, this bias may percolate to downstream tasks leading to erroneous social conclusions when applied to large-scale data for other social inference tasks. 

We use cloze tests to evaluate this implicit bias. A brief description of cloze test  follows.\\
\noindent\textbf{Cloze test:} When presented with a sentence (or a sentence stem) with a missing word, a cloze task~\cite{taylor1953cloze} is essentially a fill-in-the-blank task. For instance, in the following cloze task: \emph{In the} \texttt{[MASK]}\emph{, it snows a lot}, \texttt{winter} is a likely completion for the missing word. Word prediction as a test of LLM's language understanding has been explored in~\cite{paperno-etal-2016-lambada,ettinger-2020-bert}.  %Recent studies leveraged it in novel applications such as relation extraction~\cite{petroni-etal-2019-language} and socio-political insight mining~\cite{ecaielection2020}. 
%and fixing unsafe automatic speech recognition errors in videos meant for kids~\cite{YouTubeASR}. 

\noindent\textbf{Bias Evaluation Framework:}
We describe our proposed testing framework for gender bias. Let \texttt{LLM}$_\mathit{cloze} (w, \mathcal{S})$ denote the completion probability of the word $w$ with a masked cloze task $\mathcal{S}$ as input. For a given verb $v$, we consider the following four cloze tests: \begin{compactenum}
\item A [MASK] $v$ a woman (denoted by $v_{\textit{womanAsTheme}}$)
\item A [MASK] $v$ a man (denoted by $v_{\textit{manAsTheme}}$)
\item A man $v$ a [MASK] (denoted by $v_{\textit{manAsAgent}}$)
\item A woman $v$ a [MASK] (denoted by $v_{\textit{womanAsAgent}}$)
\end{compactenum}
In an ideal world where the LLM treats men and women equally, \texttt{LLM}$_\mathit{cloze} (\textit{man}, v_{\textit{womanAsTheme}})$ and \texttt{LLM}$_\mathit{cloze} (\textit{woman}, v_{\textit{manAsTheme}})$ should be equal. However, our preliminary exploratory analysis indicates that is not the case. For example, when $v$ is set to \textit{inspire}, \texttt{BERT}$_\mathit{cloze} (\textit{man}, v_{\textit{womanAsTheme}})$ is 0.20 whereas \texttt{BERT}$_\mathit{cloze} (\textit{woman}, v_{\textit{manAsTheme}})$ is 0.16. When we set $v$ to \textit{guide}, the gap widens -- \texttt{BERT}$_\mathit{cloze} (\textit{man}, v_{\textit{womanAsTheme}})$ is 0.71 whereas \texttt{BERT}$_\mathit{cloze} (\textit{woman}, v_{\textit{manAsTheme}})$ is 0.36.  

Again, in an ideal world where the LLM treats men and women equally, \texttt{LLM}$_\mathit{cloze} (\textit{man}, v_{\textit{womanAsAgent}})$ and \texttt{LLM}$_\mathit{cloze} (\textit{woman}, v_{\textit{manAsAgent}})$ should be equal.

Let $\mathcal{V}$ denote the set of all verbs listed in~\cite{SAPPowerAgency} where the agent has more power than the theme. Our overall measures of implicit bias are:
(a)~$(1/|\mathcal{V}|) \cdot \left( \sum_{v \in \mathcal{V} } (\texttt{LLM}_\mathit{cloze} (\textit{man}, v_{\textit{womanAsTheme}}) - \right.$ $\left. \texttt{LLM}_\mathit{cloze} (\textit{woman}, v_{\textit{manAsTheme}})) \right)$, and (b)~$(1/|\mathcal{V}|) \cdot \left(\sum_{v \in \mathcal{V} } (\texttt{LLM}_\mathit{cloze} (\textit{man}, v_{\textit{womanAsAgent}}) - \right.$ $\left. \texttt{LLM}_\mathit{cloze} (\textit{woman}, v_{\textit{womanAsAgent}})) \right)$.
% \\
% \begin{equation}
% \label{eq:Theme}
%     \frac{\sum_{v \in \mathcal{V} } (\texttt{LLM}_\mathit{cloze} (\textit{man}, v_{\textit{womanAsTheme}}) - \texttt{LLM}_\mathit{cloze} (\textit{woman}, v_{\textit{manAsTheme}}))}{|\mathcal{V}|}
% \end{equation}

% \begin{equation}
% \label{eq:Agent}
%     \frac{\sum_{v \in \mathcal{V} } (\texttt{LLM}_\mathit{cloze} (\textit{man}, v_{\textit{womanAsAgent}}) - \texttt{LLM}_\mathit{cloze} (\textit{woman}, v_{\textit{womanAsAgent}}))}{|\mathcal{V}|}
% \end{equation}

% Equation~\ref{eq:Theme} 
Measure~(a) quantifies $\textit{bias}_\textit{agent}$. A positive value indicates that the LLM encodes a man being in the position of agent likelier than a woman on expectation. 
% Equation~\ref{eq:Agent} 
Measure~(b) quantifies $\textit{bias}_\textit{theme}$. A positive value indicates that the LLM encodes a man being in the position of theme likelier than a woman on expectation.  We investigate three well-known LLMs for this audit: \texttt{BERT}~\cite{devlin2018bert}; \texttt{RoBERTa}~\cite{Roberta}; and \texttt{Megatron}~\cite{Megatron}. We consider 1,222 verbs listed in~\cite{SAPPowerAgency}. We also consider verbs in $\mathcal{X}_\textit{unpleasant}$ for this study. 

Table~\ref{tab:ClozeResults} summarizes our gender bias audit of LLMs with respect to verbs implying more power to the agent than the theme. We first note that for both verb sets, $\textit{bias}_\textit{agent}$ is substantially larger than $\textit{bias}_\textit{theme}$. This result indicates that men are considerably more likely to be considered as the agent when women is the theme and the verb implies that the agent has greater power than 
the theme. We also note that the completions favor mildly men over women even for the theme, however, the values are closer to 0. 

\begin{table*}
  % \scriptsize
  \centering
  \begin{tabular}{|l|c|c|c|c|}
    \hline
    LLM & $\mathcal{V}$, $\textit{bias}_\textit{agent}$ & $\mathcal{V}$, $\textit{bias}_\textit{theme}$ & $\mathcal{X}_\textit{unpleasant}$, $\textit{bias}_\textit{agent}$ & $\mathcal{X}_\textit{unpleasant}$, $\textit{bias}_\textit{theme}$ \TBstrut\\
    \hline
 \texttt{BERT}~\cite{devlin2018bert} & 0.32  & 0.01 & 0.23 & -0.04 
 
    \TBstrut\\
        \hline
 
    \texttt{RoBERTa}~\cite{Roberta} & 0.33  & 0.04  & 0.49  & 0.12 
 
    \TBstrut\\
   \hline 
        \texttt{Megatron}~\cite{Megatron}         & 0.30      &  0.03  & 0.34    & 0.08  
          \TBstrut\\
    \hline
         \end{tabular}

   \caption{Implicit bias in agent and theme in LLMs. $\mathcal{V}$ denotes the set of 1,222 verbs present in~Sap \textit{et el.}~\cite{SAPPowerAgency} where the agent is identified to have more power than the theme. $\mathcal{X}_\textit{unpleasant}$ denotes the set of ten unpleasant verbs considered in our study.}
    \label{tab:ClozeResults}

\end{table*}

%We lay out a blue-print for a systematic study of such biases along with preliminary evidence indicating why such biases need to be factored in for NLI tasks.

\subsection{Implicit Bias in NLI Systems}\label{Sec:BiasNLI}

We describe our approach to quantify model bias in our NLI framework specific to our task.
Consider we modify the sub-corpus $\mathcal{D}_\textit{torture}$ to $\mathcal{D}_\textit{torture}^\textit{flipped}$ where the gender identifiers in each premise sentence are flipped to the equivalent identifier of the opposite gender. For instance, the premise \textit{The \colorbox{blue!25}{wife} tortured the \colorbox{red!25}{husband} both mentally and physically} will be modified as \textit{The \colorbox{red!25}{husband} tortured the \colorbox{blue!25}{wife} both mentally and physically}. Flipping gendered words to test bias through counterfactuals in the context of coreference resolution has been previously explored in~\cite{lu2020gender}. We argue that if a premise in $\mathcal{D}_\textit{torture}$ entails \textit{A \colorbox{red!25}{man} tortures a \colorbox{blue!25}{woman}}, the flipped premise in $\mathcal{D}_\textit{torture}^\textit{flipped}$  should entail \emph{A \colorbox{blue!25}{woman} tortures a \colorbox{red!25}{man}} instead in a gender-neutral NLI system. Hence the entailment gap for \texttt{torture} computed on $\mathcal{D}_\textit{torture}$ should be equal in magnitude and opposite in polarity as the entailment gap computed on $\mathcal{D}_\textit{torture}^\textit{flipped}$. The NLI system's ($\mathcal{M}$) overall bias score with respect to verbs present in $\mathcal{X}_\textit{unpleasant}$ is thus computed as
% \begin{equation}
$NLI_\textit{bias}(\mathcal{M}, \mathcal{X}_\textit{unpleasant}) = \sum_{v \in \mathcal{X}_\textit{unpleasant}}\frac{\textit{abs}( (\textit{gap}(\mathcal{D}_v, v) + \textit{gap}(\mathcal{D}_v^\textit{flipped}, v))}{|\mathcal{X}_\textit{unpleasant}|}$.
 % \end{equation}
 In simple words, for each verb, we compute the entailment gap ($\textit{value}_1$) for the relevant sub-corpus and the flipped sub-corpus ($\textit{value}_2$). We subtract $\textit{value}_2$ from $\textit{value}_1$ and take the absolute value of the sum. The bias score is the average value of this sum across all verbs: a score close to 0 indicates that the NLI system has a minimal bias, whereas larger values indicate greater bias. 
 
 Our baseline is an off-the-shelf NLI system from Allen NLP  trained using \texttt{RoBERTa} (denoted by $\mathcal{M}_\textit{base}$). We find that $NLI_\textit{bias}(\mathcal{M}_\textit{base}, \mathcal{X}_\textit{unpleasant})$ is 0.27~\footnote{We note that a bias-aware NLI variant from Allen NLP has a better starting point (bias score 0.20) than the base model. However, the bias-aware model exhibits slower convergence than the base model when we conduct our active learning steps as discussed in Section 7.3. With identical experimental setting, after iteration 3, the bias-aware model improves its bias score to 0.133.}.

\subsection{Bias Mitigation Via Inconsistency Sampling}

\emph{Active Learning} is a powerful and well-established form of supervised machine learning technique~\cite{settles2009active} characterized by the interaction between the learner, aka the classifier, and the teacher (oracle or annotator). Each interaction step consists of the learner requesting the teacher the label of an unlabeled instance sampled using a given sampling strategy and augmenting the data set with the newly acquired label. Next, the classifier is retrained on the augmented data set. This sequential label-requesting and re-training process continues until some halting condition is reached (e.g., exceeded annotation budget or the desired classifier performance). At this point, the algorithm outputs a classifier, and the objective for this classifier is to closely approximate the (unknown) target concept in the future. The key goal of active learning is to reach a strong performance at the cost of fewer labels.

\begin{table}[htb]
\begin{center}
     \begin{tabular}{|l |c |}
    \hline
  Data & $NLI_\textit{bias}(\mathcal{M}, \mathcal{X}_\textit{unpleasant})$  \\
    \hline
  $\mathcal{M}_\textit{base}$  &  0.269     \\
    \hline 
 Iteration 1 &  0.177 $\pm$ 0.021
 \\
    \hline 
Iteration 2   &   0.110 $\pm$ 0.024
  \\ 
    \hline
Iteration 3 &    0.103 $\pm$ 0.023
  \\ 
    \hline
    \end{tabular}
    
\end{center}
\caption{Bias evaluation. Each iteration performs one round of inconsistency sampling and adds 60 samples to the train set. Performance is reported on five runs with different random seeds.}
\label{tab:baselines}

\end{table}

\begin{figure}[htb]
\centering
    \includegraphics[width=0.5\linewidth]{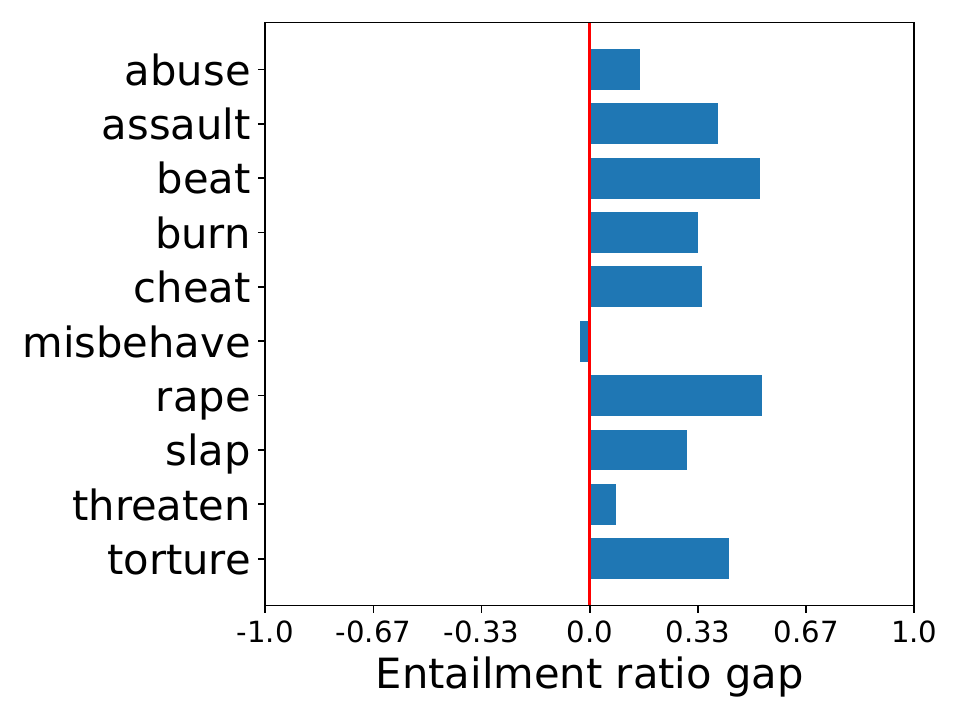}
    \caption{Gender inequality using text entailment. For a given unpleasant verb, a negative value indicates that a female has played the role of a victim more often than a male.}
 \label{fig:entailment}
\end{figure}

Some of the well-known  sampling methods include uncertainty sampling~\cite{settles2009active},  certainty sampling~\cite{sindhwani2009uncertainty}, and density-based sampling~\cite{nguyen2004active}. 
Beyond a static strategy, more complex strategies such as adapting strategy selection parameters based on estimated future residual error reduction or combining multiple sampling strategies to balance the label distribution in the procured data set have been explored in~\cite{donmez2007dual} and \cite{palakodety2020voice}, respectively.

\noindent\textbf{Inconsistency Sampling.} First introduced in Dutta \textit{et al.}~\cite{DuttaPolice}, this sampling technique exploits the underlying logical structure of the $\langle premise, hypothesis \rangle$ space. For instance, a premise cannot both entail (or contradict) a given hypothesis and its negation. In our work, we extend this idea and exploit a $\langle premise, hypothesis \rangle$ space richer than~Dutta \textit{et al.}~\cite{DuttaPolice} for logical inconsistency.

Consider the premise/hypothesis pair \textit{Continuously \colorbox{blue!25}{her} \colorbox{red!25}{husband} used to harass and torture \colorbox{blue!25}{her} everyday}/\textit{A \colorbox{red!25}{man} tortures a \colorbox{blue!25}{woman}}. We argue that if this premise entails the hypothesis (which it does), the modified premise/hypothesis pair  with replacing every gendered word with the opposite gender -- i.e., \textit{Continuously \colorbox{red!25}{his} \colorbox{blue!25}{wife} used to harass and torture \colorbox{red!25}{him} everyday}/\textit{A \colorbox{blue!25}{woman} tortures a \colorbox{red!25}{man}} -- should also entail. If not, it signals a logical inconsistency. For each sampling iteration, we add 60 samples giving equal weightage to the verbs present   in $\mathcal{X}_\textit{unpleasant}$.

Table~\ref{tab:baselines} summarizes our active learning results. For both models, $\mathcal{M}_\textit{base}$ and $\mathcal{M}_\textit{bias-aware}$, we conduct three rounds of active learning using inconsistency sampling and stop when the performance improvement becomes indiscernible ($\le 0.01$). All annotations are independently conducted by two annotators. Since legal documents are typically written in clear, unambiguous language, we observe a near-perfect agreement (Cohen's $\kappa$ value 0.96). The remaining disagreements are resolved through a post-annotation adjudication step. Table~\ref{tab:baselines} indicates that with subsequent active learning steps, our NLI system  exhibits lesser bias. Given that the maximum possible bias score is 2, we achieve substantial improvement in mitigating the bias.      

Now that we are more confident that our model inferences are  less sensitive to gender, we evaluate the societal bias present in our corpus.  
Figure~\ref{fig:entailment} summarizes our text entailment results. Barring \texttt{misbehave}, for all other verbs, men are identified as perpetrators more often than women. We further note that verbs that indicate physical abuse, such as \texttt{rape} and \texttt{beat}, particularly stand out with larger values. The average entailment gap for verbs unambiguously indicating physical harm -- \texttt{assault}, \texttt{beat}, \texttt{burn}, \texttt{slap}, and \texttt{rape} -- is much higher (0.41) than verbs that may or may not indicate physical harm (0.19) such as \texttt{abuse}, \texttt{cheat}, \texttt{misbehave}, \texttt{threaten}, and \texttt{torture}. A manual inspection of randomly sampled 200 $\langle premise, hypothesis\rangle$ pairs aligns with our automated method's overall findings.

\section{Discussions and Limitations}

In this paper, we present the first-ever computational analysis (to our knowledge) of gender inequality in divorce court proceedings in India. Based on the documented allegations of parties involved in the divorce, our analyses indicate a striking gender inequality as described in these public records.  While documented evidence of marital distress in India exists in social science literature, how such factors play out in divorce has limited understanding. Our study  sheds light on a vulnerable and vulnerable and practically invisible community in India. 

Methodologically, we identify and address several gaps and limitations of existing NLP techniques to quantify gender inequality. We believe our finding specific to legal text is new, and our method to address it is simple, effective, and intuitive. Casting the problem of quantifying gender inequality as a text entailment task is also new. Our results on text entailment results suggest that NLI can be a viable tool to computational social science researchers to analyze similar research questions (e.g., who gets the child custody can be estimated with hypotheses \emph{the husband gets the custody of the child} and \emph{the wife gets the custody of the child}). Moreover, our bias mitigation strategy exploiting a novel inconsistency sampling technique using counterfactuals holds promise. 

Our work has the following limitations.

\noindent\textbf{Sentence level processing:} An important point to keep in mind, however, is that our analyses operate at the sentence level. If in a court proceeding, a sentence records that the plaintiff accuses the defendant of wrongdoing which the defendant denies in a subsequent sentence, how these two contradicting claims are resolved in the court cannot be inferred without language models that can handle document-level contexts. We believe our research will open the gates  for investigation with newer-age LLMs that can handle broader contexts.     

%\noindent\textbf{Granular analysis based on caste, religion, and geographic regions:} India is a pluralistic nation with many religions, castes, and geographic variation. We believe our research will open the gates for future research analyzing how gender inequality manifests in divorce court proceedings along these different dimensions and can also facilitate contrastive gender inequality analyses in other cultures.

\noindent\textbf{Archival limitation:} The sparse presence of the North-Eastern region in our dataset is most likely due to archival limitation as some of these states record the highest rate of divorce~\cite{jacob2016marriage}.  Our study is also limited by the overall archival extent of \ik. 

\noindent\textbf{Economic independence:} Some of the court proceedings mention the litigants' occupations. We annotated randomly 100 sampled occupations for women.
% (see, Figure~\ref{fig:occupation} in SI). 
While an overwhelming majority of the sampled occupations are homemakers, compared to World Bank Data on labor force participation of women in India (23\%), 32\% of the women are working women in our sampled occupations. 
Economic independence and divorce merit a deeper exploration.

\noindent\textbf{Out-of-court settlements, separation, abandonment:} Finally, not all unhappy marriages end up in divorce and reach court for dissolution. Many out-of-court settlements happen. As documented in~\cite{jacob2016marriage}, the number of separated women in 2011 is almost three times the number of divorced women. Since divorce is still looked at as a social stigma~\cite{belliappa2013gender} and family institutions are highly valued in India, there could be many women who continue with their dysfunctional marriages while unhappy. The court does not know their stories.

\section{Ethical Statement}

We work with public court records. Prior studies exist on Indian court proceedings~\cite{ash2021group}. We conduct aggregate analysis refraining from presenting any personally identifiable information in the paper. Hence, we do not see any ethical concern. Rather, we believe our findings and methods can be valuable to policymakers and social scientists.  

A study on binary gender inequality runs the risk of oversimplifying gender, which we acknowledge lies on a spectrum. Same-sex marriage is yet not legal in India. Further nuances will be needed to extend our work to other cultures allowing same-sex marriages. We are also sensitive to previous studies that point out the potential harms of the erasure of gender and sexual minorities~\cite{ArjunErasurePaper}.

%\bibliographystyle{unsrt} %\bibliography{references}

% \clearpage
% \begin{center}
% \Large{Appendix}
% \end{center}

\end{document}